%% file: emnlp2021.tex
\pdfoutput=1

\documentclass[11pt]{article}

\input{preamble}

\input{definitions}

\title{\fairseqss: A Scalable and Integrable Speech Synthesis Toolkit}

\author{Changhan Wang$^{\star}$, Wei-Ning Hsu$^{\star}$, Yossi Adi, Adam Polyak, Ann Lee, \\
    \textbf{Peng-Jen Chen, Jiatao Gu, Juan Pino} \\
  Facebook AI \\
  \texttt{\{changhan,wnhsu,adiyoss,adampolyak,annl,}\\
  \texttt{pipibjc,jgu,juancarabina\}@fb.com}\\
  }

\begin{document}
\maketitle

\renewcommand{\thefootnote}{$^{\star}$}
\footnotetext[1]{Equal contribution.}
\renewcommand{\thefootnote}{1}

\begin{abstract}
This paper presents \fairseqss, a \fairseq \ extension for speech synthesis. We implement a number of autoregressive (AR) and non-AR text-to-speech models, and their multi-speaker variants. To enable training speech synthesis models with less curated data, a number of preprocessing tools are built and their importance is shown empirically. To facilitate faster iteration of development and analysis, a suite of automatic metrics is included. Apart from the features added specifically for this extension, \fairseqss~also benefits from the scalability offered by \fairseq \ and can be easily integrated with other state-of-the-art systems provided in this framework.
The code, documentation, and pre-trained models are available at \url{https://github.com/pytorch/fairseq/tree/master/examples/speech_synthesis}.
\end{abstract}

\section{Introduction}
Speech synthesis is the task of generating speech waveforms with desired characteristics, including but not limited to textual content~\citep{hunt1996unit,zen2009statistical,shen2018natural,ping2017deep,li2019neural}, speaker identity~\citep{jia2018transfer,cooper2020zero}, and speaking styles~\citep{wang2018style,skerry2018towards,akuzawa2018expressive,hsu2018hierarchical}. It is also more often referred to as \ac{TTS} when text is used as input to the system. Along with automatic speech recognition (ASR) and machine translation (MT), these language technologies have advanced rapidly over the past few years~\citep{tan2021survay}.
Traditionally, these tasks may be used in conjunction to form a system (e.g., combining the three for speech-to-speech translation), but they rarely leverage each other during training. As a result, each application used to have its own dedicated open-source toolkit, for example, Kaldi~\citep{povey2011kaldi} and HTK~\citep{young2002htk} for ASR, HTS~\citep{zen2007hmm}, Merlin~\citep{wu2016merlin}, STRAIGHT~\citep{kawahara1999restructuring}, and WORLD~\citep{morise2016world} for speech synthesis, and Moses~\citep{koehn2007moses} for MT.

Recently, there are growing interactions among these systems in the learning process. For example, \citet{hayashi2018back} and \citet{rosenberg2019speech} propose to leverage speech synthesis systems to generate paired text and speech data for ASR training; \citet{tjandra2017listening}, \citet{hori2019cycle}, and \citet{baskar2019semi} chain ASR and TTS together to form a loop for semi-supervised learning with cycle-consistency loss; \citet{weiss2017sequence}, \citet{li2020multilingual}, and \citet{jia2019direct} demonstrate that it is possible to build an end-to-end system translating speech into text or speech in a target language.

Beyond text-based systems, there is also an emerging research topic that explores the use of units discovered from self-supervised speech representation learning~\citep{oord2017neural,baevski2019vq,harwath2019learning,hsu2021hubert} to replace text for representing the lexical content in numerous applications, such as language modeling~\citep{lakhotia2021generative}, speech resynthesis~\citep{polyak2021speech}, image captioning~\citep{hsu2020text}, and translation~\citep{tjandra2020speech,hayashi2020discretalk}. This line of research bypasses the need for text and makes technologies applicable even to unwritten languages. However, to interpret the output of such systems - a sequence of learned units, a unit-to-speech model is required. This brings up the need of a framework for broader speech synthesis systems that can alternatively take learned units as input. 
These research directions can benefit from having a single toolkit with different state-of-the-art language technologies.

In this paper, we introduce \fairseqss, a \fairseq \ \citep{ott2019fairseq} extension for speech synthesis. \fairseq \ is a popular open-source sequence modeling toolkit based on PyTorch~\citep{paszke2019pytorch} that allows researchers and developers to train custom models. 
It offers great support for training large models on large scale data, and
provides a number of state-of-the-art models for language technologies. 
We extend \fairseq \ to support speech synthesis in this work. In particular, we implement a number of popular text-to-spectrogram models, with interface to both signal processing-based and neural vocoders.
Multi-speaker variants of those models are also implemented.
While speech synthesis often relies on subjective metrics such as mean opinion scores for benchmarking, we implemented a suite of widely used automatic evaluation metrics to facilitate faster iteration on model development.
Last but not least, we support a number of text and audio preprocessing modules,
which allow developers to quickly build a new dataset from less curated in-the-wild data for speech synthesis.

The main contribution of this work is threefold. First, we implement a number of state-of-the-art models and provide pre-trained checkpoints and recipes, which can be used by researchers as baselines or as building blocks in applications such as text-to-speech translation. Second, we create pre-processing tools that enable developers to use customized data to build a TTS model, and demonstrate the effectiveness of these tools empirically. Lastly, as part of the \fairseq \ codebase, this speech synthesis extension allows easy integration with numerous state-of-the-art MT, ASR, ST, LM, and self-supervised systems already built on \fairseq. We provide an example by building a unit-to-speech system that can be used for text-free research.

The rest of the paper is organized as follows: Section 2 describes the features of \fairseqss. Experiments are presented in Section 3. Related work is discussed in Section 4, and we conclude this work in Section 5.

\section{Features}
\paragraph{Fairseq Models} \fairseq~provides a collection of MT~\citep{ng-etal-2019-facebook}, ST~\citep{wang2020fairseq}, unsupervised speech pre-training and ASR~\citep{NEURIPS2020_92d1e1eb,hsu2021hubert} models that demonstrate state-of-the-art performance on standard benchmarks. They are open-sourced with pre-trained checkpoints and can be integrated or extended easily for other tasks.

\paragraph{Speech Synthesis Extension} \fairseqss~adds state-of-the-art text-to-spectrogram prediction models, Tacotron 2~\citep{shen2018natural} and Transformer~\citep{li2019neural}, which are AR with encoder-decoder model architecture. For the latest advancements on fast non-AR modeling, we provide FastSpeech 2~\citep{ren2019fastspeech,ren2020fastspeech} as an example. 
All our models support the multi-speaker setting via pre-trained~\citep{jia2018transfer} or jointly trained speaker embeddings~\citep{arik2017deep,chen2020multispeech}. Note that the former enables synthesizing speech for speakers unseen during training. For FastSpeech 2, pitch and speed are controllable during inference.
For spectrogram-to-waveform conversion (vocoding), \fairseqss~has a built-in Griffin-Lim~\citep{griffin1984signal} vocoder for fast model-free generation. It also provides examples for using external model-based vocoders, such as WaveGlow~\citep{prenger2019waveglow} and HiFiGAN~\citep{kong2020hifigan}.

\paragraph{Speech Preprocessing.} Recent advances in neural generative models have demonstrated that neural-based \ac{TTS} models, can synthesize high-quality, natural and intelligible speech. However, such models usually require high-quality, and clean speech data~\cite{zhang2021denoispeech}. In order to enable leveraging noisy data for \ac{TTS} training, we propose a speech preprocessing pipeline to enhance and filter data. The proposed pipeline is comprised of three main components: i)  Background noise removal, ii) \ac{VAD}, and iii) Outlier filtering using both \ac{SNR} and \ac{CER}. 

First, a speech enhancement model is applied over input recordings to remove background noise. We used the speech enhancement model proposed by~\cite{defossez2020real} where the $i_{th}$ convolutional layer has $2^{i-1}*64$ output channels. As suggested by the authors, we additionally used a dry/wet knob, i.e. the final output is $dry \cdot \x + (1-dry) \cdot \vyh$, where $\x$ is the noisy input signal and $\vyh$ is the output of the enhancement model. We experiment with $dry \in \{0.0, 0.01, 0.05, 0.1\}$ and find 0.01 to perform the best. 

Next, we apply \ac{VAD} to remove silence from the denoised utterances, as silence can vary in length significantly which causes increasing uncertainty and therefore degrades \ac{TTS} performance. Silence regions at the beginning and end of the utterances are completely removed. In case we encounter a silence segment in the middle of the signal in where its length is greater than 300ms we replace it with a 300ms artificially generated silence (since completely removing silence regions produces unnatural speech). Silence regions of less than 300ms are left unchanged. We use the open-source implementation of the Google WebRTC \ac{VAD}~\citep{vad}, of which four aggressiveness levels \{0, 1, 2, 3\} can be set. A higher aggressiveness level removes more silences but comes at the risk of removing partial speech. The aggressiveness level corresponds to the size of the processing window (a larger processing window will make the \ac{VAD} work at a coarser level and remove silence frames more aggressively).


Lastly, we notice that in extremely noisy recordings (\ac{SNR} close to zero), the generated denoised samples are often not intelligible enough to train a \ac{TTS} or contain distortion artifacts. In addition, when setting the VAD aggressiveness level high, speech may be truncated along with silence. To remedy this, we proposed two outliers filtering methods. The first approach is based on \ac{SNR} estimation. We approximate the noise by subtracting the output of the enhancement model from the input-noisy speech, then we compute the \ac{SNR} between the two. The second approach is based on applying an \ac{ASR} over the denoised speech and compute the CER against the target transcription. 


\paragraph{Computation} \fairseq~is implemented in PyTorch~\citep{paszke2019pytorch} and provides efficient batching, gradient accumulation, mixed precision training \citep{micikevicius2017mixed}, model parallelism, multi-GPU as well as multi-machine training for computational efficiency on large-scale experiments and enabling training gigantic models.

\paragraph{Quantitative Metrics} We provide automatic metrics for fast evaluation in model development. Similarly to~\cite{polyak2020unsupervised}, we report \ac{GPE}~\cite{nakatani2008method}, \ac{VDE}~\cite{nakatani2008method}, and \ac{FFE}~\cite{chu2009reducing} to evaluate F0 reconstructions of the generated speech. We additionally, report \ac{MCD}, \ac{MSD}, and \ac{CER} to evaluate both the overall similarity to the target speech and content intelligibility~\cite{weiss2021wave}.

\paragraph{(i) \ac{GPE}} GPE is an objective metric which measures the portion of voiced audio frames with a pitch error of more than 20\%.
\begin{equation}
    \begin{split}
    \text{GPE}&(\vp, \vph, \vv, \vvh) = \\ &\frac{\sum_t \indicator[|\vp_t - \vph_t| > 0.2 \cdot\vp_t] \indicator[\vv_t] \indicator[\vvh_t] }{\sum_t \indicator[\vv_t] \indicator[\vv'_t]}
    \end{split}
\end{equation}
where $\vp_t, \vp'_t$ are the pitch frames from the target and generated signals, $\vv_t, \vvh_t$ are the voicing decisions from the target and generated signals, and $\indicator$ is the indicator function. 

\paragraph{(ii) \ac{VDE}} VDE measures the portion of frames with voicing decision error,
\begin{equation}
    \text{VDE}(\vv, \vvh) = \frac{\sum_{t=1}^{T-1} \mathbbm{1}[\vv_t \ne \vvh_t]}{T},
\end{equation}
where $T$ is the total number of frames.

\paragraph{(iii) FFE} Combining GPE and VDE, FFE measures the percentage of frames that contain a deviation of more than 20\% in pitch value or have a voicing decision error.
\begin{equation}
    \begin{split}
    \text{FFE}&(\vp, \vph, \vv, \vvh) = \text{VDE}(\vv, \vvh) \\ + &\frac{\sum_{t=1}^{T-1} \indicator[|\vp_t - \vph_t| > 0.2\cdot\vp_t] \indicator[\vv_t] \mathbbm{1}[\vvh_t]}{T}.
    \end{split}
\end{equation}

\paragraph{(iv) \ac{MCD}/\ac{MSD}} These are defined as the root mean squared error of the synthesized speech against the reference speech computed on the 13-dimensional MFCC features for \ac{MCD} and log-mel spectral features for MSD. Since the reference and the synthesized speech may not be aligned frame-by-frame, instead of zero-padding the shorter one and assuming they are frame-wise aligned as done in \citet{skerry2018towards}, we follow \citet{weiss2021wave} and use dynamic time warping~\citep{berndt1994using} to align the frames from the two sequences. The main difference between these two metrics lies in the features they compute distortion on: MFCC features aim to capture phonetic information while removing speaker information, while log-mel spectral features encode both, and hence \ac{MCD} addresses phonetic similarity more.

\paragraph{(v) \ac{CER}} CER is computed between the transcription of the generated audio against the input text using an \ac{ASR} system publicly available in \fairseq. 

\paragraph{Visualization} \fairseq~integrates Tensorboard\footnote{\url{https://github.com/tensorflow/tensorboard}} for monitoring holistic metrics during model training. It also has VizSeq~\citep{wang-etal-2019-vizseq} integration for offline sequence-level error analysis, where transcript and target/predicted speech are visualized in Jupyter Notebook interface. \fairseqss~further adds generated spectrogram and waveform samples to Tensorboard for model debugging.

\input{table_ljspeech}
\section{Experiments}
We evaluate our models in three settings: single-speaker synthesis, multi-speaker synthesis and multi-speaker synthesis using noisy data.

\subsection{Experimental Setup}
We use either characters, phonemes or discovered units as input representations.
To convert texts into phonemes, we employ g2pE~\citep{g2pE2019} or Phonemizer~\citep{phonemizer2015} with espeak-ng\footnote{\url{https://github.com/espeak-ng/espeak-ng}} backend. We use the Montreal Forced Aligner~\citep{mcauliffe2017montreal} to obtain phonemes with frame durations for FastSpeech 2 training, which is based on the same pronunciation dictionary (CMUdict) as g2pE. 
For discovered units, we extract frame-level units using a Base HuBERT model trained on LibriSpeech\footnote{\url{https://dl.fbaipublicfiles.com/hubert/hubert_base_ls960.pt}} and collapse consecutive units of the same kind. We use the run length of identical units before collapsing as target duration for FastSpeech 2 training. We use a reduction factor (number of frames each decoder step predicts) of 4 for Transformer and 1 for FastSpeech 2 by default.

We resample audios to 22,050Hz and extract log-Mel spectrogram with FFT size 1024, window length 1024 and hop length 256. We optionally pre-process audios to improve model training: denoising (``DN"), level-2 or level-3 VAD (``VAD-2" or "VAD-3"), filtering by SNR$>15$ and CER$<10\%$ (``FLT") and volume normalization (``VN").

We use MCD and CER for automatic evaluation. MCD is computed on Griffin-Lim vocoded reference and model output spectrograms. We use vocoded references as opposed to the original ones to eliminate the error introduced by the vocoder and focus the evaluation on spectrogram prediction.
HiFiGAN vocoders trained on each dataset are used to generate waveforms for CER evaluation. 
The large wav2vec 2.0~\cite{baevski2020wav2vec} ASR model, which achieves WERs of 1.8\% and 3.3\% on Librispeech test-clean and test-other, respectively and is provided in \fairseq\footnote{\url{https://dl.fbaipublicfiles.com/fairseq/wav2vec/wav2vec_vox_960h_pl.pt}}, is used both for CER filtering and evaluation.
GPE, VDE, and FFE are not reported here, because these metrics are more meaningful when prosody modeling is taken into account~\citep{polyak2020unsupervised, skerry2018towards,wang2018style}.
For subjective evaluation, we conduct a Mean Opinion Score (MOS) test using the CrowdMOS package~\cite{ribeiro2011crowdmos} using the recommended recipes for detecting and discarding inaccurate scores. We randomly sample 100 speech utterances from the test set and collect manual scores using a crowd sourcing framework. The same samples are used across all tested methods. Each sample is rated by at least 10 raters on a scale from 1 to 5 with 1.0 point increments. Overall, scores for each tested method are averaged across more than 1000 manual annotations. We report both average MOS scores together with a 95\% confidence interval (CI95).

\subsection{Single-Speaker Synthesis on LJSpeech}
\input{table_vctk}
\input{table_common_voice}

\input{table_counterparts}

LJSpeech~\citep{ljspeech17} is a single-speaker TTS corpus with 13,100 English speech samples (around 24 hours) from audiobooks. We follow the setting in~\citet{ren2020fastspeech} to use 349 samples (with document title LJ003) for validation, 523 samples (with document title LJ001 and LJ002) for testing and the rest for training.

On this de-facto standard benchmark, we compare autoregressive model (Transformer, ``TFM") with non-autoregressive model (FastSpeech 2, ``FS2"), as well as 3 different types of inputs: characters, phonemes (from g2pE or espeak-ng) and HuBERT units. We see from Table~\ref{tab:ljspeech} that FastSpeech 2 performs comparably well to Transformer with phoneme inputs (g2pE), both achieving 4.2 MOS. However, the latter does not require input-output alignments for model training and supports more types of inputs---it achieves 4.1 MOS with characters (no need for phonemization), and 4.2 MOS with simpler phonemes (espeaker-ng). The task falls into the re-synthesis setting with unit inputs. We notice that FastSpeech 2 performs worse (4.0 vs. 4.2 on MOS) in this setting, likely due to the finer-grained inputs and its simplified attention mechanism.

\begin{figure*}[ht]
    \centering
    \includegraphics[width=\linewidth]{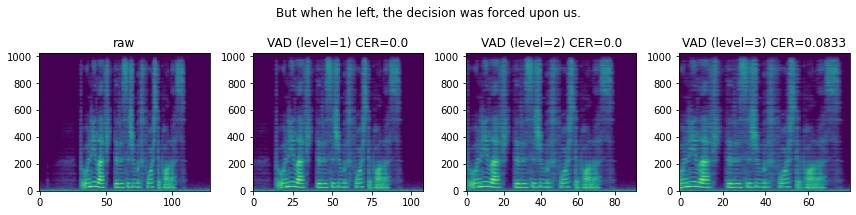}
    \caption{\textbf{A VCTK example.} With VAD level 3, the first word ``But'' is detected as silence and cut off.}
    \label{fig:vctk}
\end{figure*}

\subsection{Multi-Speaker Synthesis on VCTK}

VCTK~\citep{veaux2017superseded} is a multi-speaker English TTS dataset that contains 44 hours of read speech from 109 speakers with various English accents\footnote{\url{https://datashare.ed.ac.uk/handle/10283/3443}}. We randomly sample 50 utterances for validation and 100 utterances for testing, and use the rest for training.

Speech recordings from VCTK include considerable amount of silence as shown in \autoref{fig:vctk} (raw); therefore, silence removal is considered a standard preprocessing step for VCTK~\citep{jia2018transfer, cooper2020zero}. \autoref{fig:vctk} shows silence-removed spectrograms with three VAD aggressiveness levels. We see that a higher aggressiveness level removes more silence, but may also truncate the speech. The dataset durations after silence removal and filtering with CER $<10\%$ are listed in \autoref{tab:vctk_preproc}, along with the validation CER.

We use this dataset to study how audio-preprocessing and speaker representation affect the performance of TTS. We train a transformer TTS model with a reduction factor (i.e.\ how many frames each decoding step predicts) of 2 or 4 on three sets of audio: raw data (Raw), DN+VAD-3, and DN+VAD-3+FLT. A speaker embedding lookup table (LUT) is used by default. In addition, we train models on DN+VAD-3+FLT with a fixed embedding (Emb) for each speaker inferred from a pre-trained speaker verification model~\citep{heigold2016end}, which would enable synthesizing the voice of an unseen speaker.
Results in Table~\ref{tab:vctk} show that increasing the reduction factor from 2 to 4 improves the performance consistently.
Specifically, we found that without VAD, the model fails to train when using a reduction factor of 2. Finally, we found that using a pre-trained speaker embedder achieves similar performance than using a learnable lookup table, while enabling synthesizing speech for unseen speakers. 

\subsection{Multi-Speaker Synthesis using Noisy Data from Common Voice}

Common Voice~\citep{ardila-etal-2020-common} is a multi-speaker speech corpus with around 4.2K hours of read speech in 40 languages (version 4). It is crowd-sourced from around 78K voice contributors in various accents, age groups and genders. We use its English portion and select data from the top 200 speakers by duration (total 226 hours).

The audio data in this corpus is expectedly noisy given the lack of curated recording environments. We explore if speech processing can counteract the negative factors (background noise, long silence, variable volume across clips, etc.) during recordings and improve model training. Specifically, we examine 3 preprocessing settings with Transformer model and phoneme (g2pE) inputs: VN, DN+VAD-2+VN and DN+VAD-2+FLT+VN. As shown in Table~\ref{tab:common_voice}, the original audio has 0.3/0.5 lower MOS than the LJSpeech/VCTK one, confirming its relatively low recording quality. Noise and silence removal improve synthesis quality significantly by 0.2 MOS (DN+VAD-2+* vs. VN). Filtering by SNR and CER improves both model fitting (-0.1 MCD) and intelligibility (-1.5 CER) given the removal of difficult training examples. 

\section{Related Work}
There are many existing open-source repositories for speech synthesis. The most prominent toolkits for conventional statistical parametric speech synthesis (SPSS) include HMM/DNN-based Speech Synthesis System (HTS)~\citep{zen2007hmm} and Merlin~\citep{wu2016merlin}. These rely heavily on feature engineering and 
use signal processing-based vocoders like STRAIGHT~\cite{kawahara1999restructuring} and WORLD~\cite{morise2016world} to synthesize waveforms from acoustic features (e.g., fundamental frequency, spectral envelope, and aperiodic information).
Recently, end-to-end models that take minimally pre-processed features (characters and mel-spectrograms) have achieved superior performance compared to conventional systems~\citep{shen2018natural}, especially when paired with neural vocoders~\citep{prenger2019waveglow,kong2020hifigan}. There are a number of open-source implementations available on Github~\footnote{coqui-ai/TTS, Kyubyoung/tacotron, NVIDIA/tacotron2,  Rayhane-mamah/Tacotron2, r9y9/deepvoice3\_pytorch},
however, these repositories are solely for text-to-speech synthesis, and mostly support one model only.

ESPnet~\citep{watanabe2018espnet,hayashi2020espnet}, NeMo, and OpenSeq2Seq~\cite{kuchaiev-etal-2018-openseq2seq} are the most similar toolkits that also support multiple tasks. As listed in Table~\ref{tab:counterparts}, \fairseqss~provides more audio preprocessing tools and automatic metrics for building and evaluating speech synthesis models on custom datasets. As part of \fairseq, it can also be easily integrated with numerous state-of-the-art models already provided in \fairseq \ for exploring novel ideas. For example, we demonstrate that units discovered from a self-supervised speech pre-training model can be used to build a unit-to-speech system that converts output from systems like unit LM~\citep{lakhotia2021generative} or image-to-unit~\citep{hsu2020text} to speech.

\section{Conclusion}
This paper introduces \fairseqss, a \fairseq \ extension for speech synthesis. We believe this extension will allow researchers and developers to more easily test novel ideas for language technologies by providing great support for scalability, integrability, and a wealth of tools for curating data as well as automatically evaluating trained systems.



\bibliography{custom}
\bibliographystyle{acl_natbib}

\end{document}

%% file: preamble.tex
\usepackage[]{emnlp2021}

\usepackage{times}
\usepackage{latexsym}

\usepackage[T1]{fontenc}

\usepackage[utf8]{inputenc}

\usepackage{microtype}

\usepackage{multirow}
\usepackage{acronym}
\usepackage{hyperref}

\acrodef{VAD}{Voice Activity Detector}
\acrodef{SNR}{Signal-to-Noise Ratio}
\acrodef{ASR}{Automatic Speech Recognition}
\acrodef{TTS}{Text-to-Speech}
\acrodef{WER}{Word Error Rate}
\acrodef{CER}{Character Error Rate}
\acrodef{MCD}{Mel Cepstral Distortion}
\acrodef{MSD}{Mel Spectral Distortion}
\acrodef{GPE}{Gross Pitch Error}
\acrodef{VDE}{Voicing Decision Error}
\acrodef{FFE}{F0 Frame Error}

%% file: definitions.tex
\newcommand{\fairseq}{\textsc{fairseq}}
\newcommand{\fairseqss}{\textsc{fairseq S$^2$}}

\usepackage{bm}
\usepackage{amssymb,amsmath,graphicx,bbm,color,booktabs}
\usepackage{amsopn}

\newcommand{\vp}{\bm{p}}       \newcommand{\vph}{\hat{\bm{p}}}

\newcommand{\vv}{\bm{v}}       \newcommand{\vvh}{\hat{\bm{v}}}

       \newcommand{\vyh}{\hat{\bm{y}}}

\newcommand{\indicator}{\mathbbm{1}}

\renewcommand{\eqref}[1]{Eq.~(\ref{#1})}

\def \x{{\mathbf x}}

%% file: table_ljspeech.tex
\begin{table}[t]
    \small
    \centering
    \begin{tabular}{lr|cc|c}
    \toprule
     & & MCD & CER (S/D/I) & MOS \\
    \midrule
    \multicolumn{2}{c|}{Orig. Audio}  & - & 3.3 (0.2/0.5/2.5) & 4.53$\pm$0.05 \\
    \midrule
    \multirow{4}{*}{TFM} & Char & 4.1 & 4.4 (0.8/0.8/2.8) & 4.09$\pm$0.06 \\
    & g2pE & 3.8 & 5.0 (1.1/1.2/2.7) & 4.18$\pm$0.06 \\
    & espk & 4.4 & 3.8 (0.5/0.6/2.7) & 4.17$\pm$0.06 \\
    & Unit & 3.4 & 5.7 (1.4/1.3/3.1) & 4.18$\pm$0.05 \\
     \midrule
    \multirow{2}{*}{FS2} & g2pE & 3.8 & 4.9 (1.2/0.9/2.8) & 4.15$\pm$0.09 \\
     & Unit & 3.4 & 7.6 (2.6/1.8/3.2) & 3.99$\pm$0.05 \\
    \bottomrule
    \end{tabular}
    \caption{\textbf{Evaluation on LJSpeech.} We compare autoregressive model (``TFM") with non-autoregressive model (``FS2"), as well as 3 different types of inputs: characters (``char"), phonemes (``g2pE" and ``espk") and HuBERT units (``unit"). 
    }
    \label{tab:ljspeech}
\end{table}

%% file: table_vctk.tex
\begin{table}[t]
    \small
    \centering
    \begin{tabular}{r|cc}
    \toprule
    Audio Preprocessing & Hours & CER (dev) \\
    \midrule
    Raw & 44 & 0.8  \\
    DN+VAD-1 & 33 & 1.0 \\
    DN+VAD-2 & 32 & 1.2  \\
    DN+VAD-3 & 26 & 6.8  \\
    DN+VAD-3 + FLT & 20 & 1.6  \\
    \bottomrule
    \end{tabular}
    \caption{\textbf{Audio preprocessing settings on VCTK.} FLT removes samples with CER $>10\%$.}
    \label{tab:vctk_preproc}
\end{table}

\begin{table}[t]
    \small
    \centering
    \resizebox{\linewidth}{!}{
    \begin{tabular}{rcc|cc|c}
    \toprule
    Audio & Spk. & Red. & MCD & CER & MOS \\
    \midrule
    Original & - & - & - & 1.8 & 4.27$\pm$0.07 \\
    \midrule
    \multirow{2}{*}{Raw} & \multirow{2}{*}{LUT} 
          & 2 & 4.9 & 65.2 & 1.77$\pm$0.08 \\
        & & 4 & 3.3 & 12.1 & 2.77$\pm$0.08 \\
    \midrule
    \multirow{2}{*}{DN+VAD-3} & \multirow{2}{*}{LUT} 
          & 2 & 3.6 & 9.8 & 3.34$\pm$0.06 \\
        & & 4 & 3.4 & 6.9 & 3.30$\pm$0.06 \\
    \midrule
    & \multirow{2}{*}{LUT} & 2 & 3.6 & 9.7 & 3.38$\pm$0.06 \\
    DN+VAD-3 & & 4 & 3.4 & 6.0 & 3.42$\pm$0.05 \\
    \cmidrule{2-6}
    +FLT & \multirow{2}{*}{Emb} 
     & 2 & 3.6 & 7.6 & 3.38$\pm$0.06 \\
    & & 4 & 3.5 & 5.8 & 3.25$\pm$0.08 \\
    \bottomrule
    \end{tabular}
    }
    \caption{\textbf{Evaluation on VCTK.} We use Transformer with character inputs, and compare 3 audio pre-processing settings and 2 types of speaker embeddings.}
    \label{tab:vctk}
\end{table}

%% file: table_common_voice.tex
\begin{table}[t]
    \small
    \centering
    \begin{tabular}{r|cc|c}
    \toprule
    Audio & MCD & CER & MOS \\
    \midrule
    Original & - & 3.0 & 4.0$\pm$0.06 \\
    \midrule
    VN & 5.5 & 19.6 & 2.97$\pm$0.08 \\
    DN+VAD-2+VN & 5.6 & 9.8 & 3.22$\pm$0.07 \\
    DN+VAD-2+FLT+VN & 5.6 & 9.2 & 3.17$\pm$0.06 \\
    \bottomrule
    \end{tabular}
    \caption{\textbf{Evaluation on Common Voice English portion (top 200 speakers only).} We use Transformer model with phoneme (g2pE) inputs and compare 3 audio preprocessing settings.
    }
    \label{tab:common_voice}
\end{table}

%% file: table_counterparts.tex
    

\begin{table*}[ht]
    \centering
    \small
    \begin{tabular}{c|ccccccccc}
    \toprule
    
      & Multi-Spk & Non-AR  & \multirow{2}{*}{ASR} & \multirow{2}{*}{MT} & \multirow{2}{*}{ST} & Speech       & Audio       & Auto. \\
      & TTS       & TTS     &                      &                     &                     & Pre-training & Preprocess  & Metrics \\
    \midrule
    coqui TTS$^1$ & \checkmark & \checkmark & & & & & &  \\
    OpenSeq2seq$^{2\dagger}$ & & & \checkmark & \checkmark & & & & \\
    ESPnet-TTS$^3$ & \checkmark & \checkmark & \checkmark & \checkmark & \checkmark & & \checkmark$^\ddagger$ & \checkmark$^\ddagger$ \\
    NeMo$^4$ & \checkmark & \checkmark & \checkmark & \checkmark & & & & \\
    \midrule
    \textbf{\fairseqss} & \checkmark & \checkmark & \checkmark & \checkmark & \checkmark & \checkmark & \checkmark & \checkmark \\
    \bottomrule
    \end{tabular}
    \caption{\textbf{Comparison of \fairseqss~with counterpart speech synthesis toolkits (as of June 2021).} $^1$ GitHub: coqui-ai/TTS. $^2$ \citet{kuchaiev2018openseq2seq}. $^3$ \citet{hayashi2020espnet}. $^4$ GitHub: NVIDIA/NeMo. $^\dagger$ Archived and no longer updated. $^\ddagger$ Supporting only VAD for audio preprocessing and MCD for automatic metric.
    }
    \label{tab:counterparts}
\end{table*}